\documentclass[pra,aps,twocolumn,superscriptaddress,showpacs,showkeys]{revtex4-1}
\usepackage{amsmath,amsthm,amssymb,graphicx,subfigure,xcolor}
\usepackage{graphics,float}
\usepackage{color}
\usepackage[unicode=true]{hyperref}
\hypersetup{
     colorlinks=true,       		
     linkcolor=blue,          	
     citecolor=red,            
     urlcolor=magenta,           	
 }

\begin{document}

\title{Hardy's paradox for multi-settings and high-dimensional systems}
\author{Hui-Xian Meng}
\affiliation{Theoretical Physics Division, Chern Institute of Mathematics, Nankai University, Tianjin 300071, People's Republic of China}

\author{Jie Zhou}
\affiliation{Theoretical Physics Division, Chern Institute of Mathematics, Nankai University, Tianjin 300071, People's Republic of China}

\author{Zhen-Peng Xu}
\affiliation{Theoretical Physics Division, Chern Institute of Mathematics, Nankai University, Tianjin 300071, People's Republic of China}

\author{Hong-Yi Su}
\email{hysu@mail.nankai.edu.cn}
\affiliation{Graduate School of China Academy of Engineering Physics, Beijing 100193, People¡¯s Republic of China}

\author{Ting Gao}
\affiliation{Theoretical Physics Division, Chern Institute of Mathematics, Nankai University, Tianjin 300071, People's Republic of China}
\affiliation{College of Mathematics and Information Science,Hebei Normal University, Shijingzhuang 050024, People's Republic of China}

\author{Feng-Li Yan}
\affiliation{Theoretical Physics Division, Chern Institute of Mathematics, Nankai University, Tianjin 300071, People's Republic of China}
\affiliation{College of Physics Science and Information Engineering, Hebei Normal University, Shijiazhuang 050024, People's Republic of China}

\author{Jing-Ling Chen}
\email{chenjl@nankai.edu.cn}
\affiliation{Theoretical Physics Division, Chern Institute of Mathematics, Nankai University,
 Tianjin 300071, People's Republic of China}
\affiliation{Centre for Quantum Technologies, National University of Singapore,
 3 Science Drive 2, Singapore 117543}

\date{\today}
\begin{abstract}
Recently, Chen et al introduced an alternative form of Hardy's paradox for $2$-settings and high-dimensional systems [Phy. Rev. A 88, 062116 (2013)], in which there is
a great progress in improving
the maximum
probability of the nonlocal event.
 Here,
we construct a general Hardy's paradox for multi-settings and high-dimensional systems, which (i) includes  the paradox in [Phy. Rev. A 88, 062116 (2013)] as a special case, (ii)
 for  spin-$\frac{1}{2}$ systems,   is equivalent to the ladder proof of nonlocality without inequalities  in [Phy. Rev. Lett. 13, 2755 (1997)], (iii)
 for  spin-$1$ systems, increases the maximum
probability of the nonlocal event  by adding the number of settings, specially,
with only $5$-settings it can be improved to 0.40184, which  is more than two times higher than
 0.171, the maximal success probability to prove nonlocality in Adan's paradox [Phy. Rev. A 58, 1687 (1998)].
\end{abstract}

\pacs{03.65.Ud, 03.67.Mn, 42.50.Xa}
\keywords{}
\maketitle

\section{Introduction}

 Hardy's paradox is
the simplest  demonstration of Bell nonlocality, the impossibility of describing all quantum correlations in terms of local hidden variables \cite{Bell1964,Hardy1992,Hardy1993}. In the original form, 
the paradox occurs when some two-particle entangled states are measured by
two observers, each with two  von Neumann
measurements, each with two outcomes. However, if one wants to study  nonlocality from the point of paradoxes in a systematic
way, one must investigate a general paradox for any given number of
parties, settings, and outcomes, and check the tightness of Bell's inequalities induced by the paradox.

For two spin-half particles, increasing the number of settings  at each end is an efficient method to improve the
successful probability of demonstration of ``nonlocality without inequalities'' \cite{Hardy1997}. Subsequently, a similar proof of nonlocality  is extended to two
spin-$1$ particles \cite{Cabello1998},  and using $5$-settings at each end the proof worked for 0.171 of pairs. But, for a long time, the methods of extending Hardy's paradox to high-dimensional systems can not improve the maximal success probability \cite{Kunkri,Seshadreesan,Rabelo}.
Until 2013, from  an paradox, equivalent to Hardy's paradox for spin-$1/2$ systems, for spin-$s$ ($s\geq 1/2$) system \cite{Chen2013},  the maximal probability of the nonlocal events can grow with the dimension
of the local systems. Up to now,  a number of experiments have been carried out to
confirm the nonlocality without inequalities \cite{Torgerson1995,Giuseppe1997,Hardy1997,Lundeen 2009,Vallone 2011,Fedrizzi 2011,Chen 2012,Karimi 2014} in two-particle systems.
Hardy's paradox has also been extended to the case of more than two particles \cite{Cereceda 2012,Chen 2018}.
Hardy-like proofs can also be applied to
contextuality \cite{Cabello 2013}. Despite these fruitful achievements, Hardy's paradox has not been extended to  any number of settings and any two high-dimensional systems.

The aim of this work is to present a general Hardy's paradox for multi-settings and high-dimensional systems. It
 degenerates to the paradox in \cite{Hardy1997} for $k$-setting measurements and spin-$\frac{1}{2}$ systems and the paradox in \cite{Chen2013} for $2$-setting measurements and spin-$s$ systems.
For  spin-$1$ systems, this paradox
shows that using only $5$-settings the maximum
probability of nonlocal events can attain 0.40184 which  is almost
three times of  0.1413, the maximum
probability of nonlocal events of the Hardy's paradox in \cite{Chen2013}. Moreover,
we introduce generalized Hardy's inequalities
for any given number of settings and outcomes, and find that they are tight for $k$-settings and spin-$1$ system, where $k=3,4,5,6.$
Arguably, all these features make this paradox
of fundamental importance.

\section{Hardy's paradox for $k$-settings and $d$-dimensional systems}

A bipartite $d$-dimensional system can be described by a quantum pure state
\begin{equation}
  |\psi\rangle=\sum_{i,j}h_{ij}|i\rangle\otimes|j\rangle,~~~i,j=0,1,...,d-1,
\end{equation}
where $|i\rangle$'s are a set of orthonomal bases, and $h_{ij}$'s denote the coefficients that satisfy the normalization requirement: $\sum_{i,j} |h_{ij}|^2=1$. Below, for simplicity, it is tacitly assumed that these coefficients are all real-valued. The state can thus be represented uniquely by a coefficient matrix
\begin{equation}
  H=\left[ \begin{matrix}
    \ddots &   &  \\
           & h_{ij} &  \\
           &   & \ddots
  \end{matrix} \right].
\end{equation}

In this paper, we consider two observers: Alice, who can  make only one measurement $A_i$ from a set of
$\{A_i:i=1,2,\cdots,k\}$ on her subsystem, and Bob, who can also make only one measurement $B_j$
from a set of
$\{B_j:j=1,2,\cdots,k\}$ on his. Suppose that each of these measurements has
$d$ outcomes that we will number as $0,1,2,\cdots,d-1$. In the following, we assume that the  von Neumann measurements (VNMs) form as
$$A_i=\{|A_{i,0}\rangle\langle A_{i,0}|,|A_{i,1}\rangle\langle A_{i,1}|,\cdots,|A_{i,d-1}\rangle\langle A_{i,d-1}|\},$$
$$B_i=\{|B_{i,0}\rangle\langle B_{i,0}|,|B_{i,1}\rangle\langle B_{i,1}|,\cdots,|B_{i,d-1}\rangle\langle B_{i,d-1}|\},$$
where $$\{|A_{i,s}\rangle:s=0,1,\cdots,d-1\}$$ and $$\{|B_{i,t}\rangle:t=0,1,\cdots,d-1\}$$ are orthonormal bases of $d$-dimensional system $\mathbb{C}^d.$
Following the symbols used in \cite{Chen2013},
 $P(A_i< B_j)$ denoted the joint conditional probability that
the result of $A_i$ is strictly smaller than the result of $B_j$. Then
\begin{equation}\label{eqnp}P(A_i< B_j)=\sum_{s=0}^{d-2}\sum_{t=s+1}^{d-1}|\langle\psi|A_{i,s}\rangle|B_{j,t}\rangle|^2.
\end{equation}

Following from the fact that, according to
quantum theory, there exist two-qudit entangled states and
local measurements satisfying, simultaneously,
 \begin{equation}\label{eqH}
 \begin{split}
 &P(A_k<B_k)>0,\\
 &P(A_i<B_{i-1})=0,\\
 &P(B_{i-1}<A_{i-1})=0,\\
  &P(A_1<B_k)=0,
  \end{split}
 \end{equation}
 for any $i=2,3,\cdots,k$. However, if  events  $A_i<B_{i-1},B_{i-1}<A_{i-1},(i=2,3,\cdots,k),A_1<B_k$  never happen, then, in any local theory, the event $A_k<B_k$ never happen. For $k=2$, it is just the Hardy's paradox for $d$-dimensional systems presented by Chen in \cite{Chen2013}.
 Therefore,  (\ref{eqH}) is {\it{a general Hardy's paradox for $k$-settings and $d$-dimensional systems.}}

Let us define
\begin{equation}\label{eqmn}SP_{k,d}=\max P(A_k<B_k)\end{equation}
satisfying conditions in (\ref{eqH}). Then $SP_{k,d}$ denotes the maximal successful probability to prove nonlocality in Hardy's paradox (\ref{eqH}) for $k$-settings and $d$-dimensional systems.

 \section{Hardy's paradox (\ref{eqH}) for two-qubit systems}

For two-qubit systems, each of both Alice's and Bob's measurements  has only
$2$ outcomes, $0$ or $1$.
By the constrain conditions
$P(A_{k}<B_{k-1})=0$  and
$P(B_{k-1}<A_{k-1})=0$ in (\ref{eqH}),
 we obtain
$$\langle\psi|A_{k,0}\rangle|B_{k-1,1}\rangle=0,{\rm{i.e.,}}\ |B_{k-1,1}\rangle\bot H^T|A_{k,0}\rangle$$
 and
$$\langle\psi|A_{k-1,1}\rangle|B_{k-1,0}\rangle=0,{\rm{i.e.,}}\ |A_{k-1,1}\rangle\bot H|B_{k-1,0}\rangle$$
respectively,
which imply
$$|B_{k-1,0}\rangle\propto H^T|A_{k,0}\rangle$$
by the orthogonality of $|B_{k-1,1}\rangle$ and $|B_{k-1,0}\rangle$,
and
$$|A_{k-1,0}\rangle\propto H|B_{k-1,0}\rangle\propto HH^T|A_{k,0}\rangle,$$
by the orthogonality of $|A_{k-1,0}\rangle$ and $|A_{k-1,1}\rangle$.
Similarly, for any $i=1,2,\cdots,k,$ we can  represent $|A_{i,0}\rangle,|A_{i,1}\rangle,|B_{i,0}\rangle,|B_{i,1}\rangle$ with $|A_{k,0}\rangle$ and $H$. In particular, $$|B_{k,0}\rangle \propto H^T(HH^T)^{k-1}|A_{k,0}\rangle.$$ To calculate $SP_{k,2}$, it is sufficient to  take
$$H=\left(
      \begin{array}{cc}
        \cos{\theta} & 0 \\
        0 & \sin{\theta} \\
      \end{array}
    \right),|A_{k,0}\rangle=\left(
                              \begin{array}{c}
                                \cos{\phi} \\
                                \sin{\phi} \\
                              \end{array}
                            \right),
$$
and then $$|B_{k,0}\rangle \propto \left(
                              \begin{array}{c}
                                \cos^{2k-1}{\theta}\cos{\phi} \\
                                \sin^{2k-1}{\theta}\sin{\phi} \\
                              \end{array}
                            \right),$$
                            $$|B_{k,1}\rangle \propto \left(
                              \begin{array}{c}
                                 \sin^{2k-1}{\theta}\sin{\phi} \\
                                -\cos^{2k-1}{\theta}\cos{\phi}\\
                              \end{array}
                            \right).$$
Therefore,
\begin{eqnarray*}
& &P(A_k<B_k)\\
&=&|\langle \phi|A_{k,0}\rangle|B_{k,1}\rangle|^2\\
&=&\frac{\sin^2{\phi}\cos^2{\phi}\cos^2{\theta}\sin^2{\theta}
(\sin^{2k-2}{\theta}-
\cos^{2k-2}{\theta})^2}{(\sin^{2k-1}{\theta}\sin{\phi})^2+(\cos^{2k-1}{\theta}\cos{\phi})^2},
\end{eqnarray*}
which implies that the paradox in (\ref{eqH}) is equivalent to the ladder proof of nonlocality in \cite{Hardy1997}.
In table \ref{table1}, we list $SP_{k,d}$ for $k=2,3,4,5,6$ and $d=2$.
 As $k\rightarrow \infty$, $SP_{k,2}\rightarrow 0.5$ and  $P(A_k<B_k)=0$ for the maximally
entangled state.

\begin{table}[H]
\caption{\label{table1}For two-qubit systems, the maximal successful probability to prove nonlocality in Hardy's paradox (\ref{eqH}) for $k=2,3,4,5,6$.}
\begin{ruledtabular}
\begin{tabular}{cccccc}
 &  $k=2$ & $k=3$ & $k=4$& $k=5$ & $k=6$\\
  \hline
  $SP_{k,2}$ &  $0.09017$ & $0.17455$ & $0.23126$& $0.27088$ & $0.29995$\\
\end{tabular}
\end{ruledtabular}
\end{table}

\section{Hardy's paradox (\ref{eqH}) for two-qutrit systems}

Two-qutrit system is much richer than two-qubit system. For  $2$-setting scenarios, \cite{Chen2013} introduced alternative formulation of Hardy's paradox, which is just the paradox (\ref{eqH}) with $k=2$, and
numerically proved that  $SP_{2,3}=P_{{\rm{Hardy}}}^{d=3}\approx 0.1413$ the maximal probability of nonlocal events can be  higher than $SP_{2,2}=P_{{\rm{Hardy}}}^{d=2}\approx 0.0917$.
 Next, we
 investigate $SP_{k,3}$ for Hardy's paradox (\ref{eqH}) as follows:

\begin{itemize}
  \item $k$=3: It is sufficient to let
 Alice and Bob choose VNMs:
$$A_1=B_1=\{|0\rangle\langle 0|,|1\rangle\langle 1|,|2\rangle\langle 2|\},$$
then condition $P(B_1<A_1)=0$ leads to $h_{ij}=0$ for $i>j$. This implies that the matrix $H$ is
an upper-triangular matrix. Condition $P(A_1<B_3)=0$ leads to
$$\langle B_{3,2}| \bot \langle 0| H,\langle B_{3,2}| \bot \langle 1| H,\langle B_{3,1}| \bot \langle 0| H, $$
because of  mutual orthogonality of $\langle B_{3,2}|,\langle B_{3,1}|$ and $\langle B_{3,0}|$, we can use entries of $H$ to denote the VNM $B_3$. Condition $P(A_2<B_1)=0$ implies
$$|A_{2,0}\rangle\bot H |1 \rangle,|A_{2,0}\rangle\bot H |2\rangle,|A_{2,1}\rangle\bot H |2\rangle,$$
because of  mutual orthogonality of $|A_{2,0}\rangle,|A_{2,1}\rangle$ and $|A_{2,2}\rangle$, we can use entries of $H$ to denote the VNM $A_2$. Similarly, by the constraint conditions in Hardy's paradox (\ref{eqH}), we can also use the elements of $H$ to denote VNMs $B_2$ and $A_3$.
Thus,
  \begin{eqnarray*}
  ~~~~~~P(A_3<B_3)&=&|\langle \phi|A_{3,0}\rangle|B_{3,1}\rangle|^2+
 |\langle \phi|A_{3,0}\rangle|B_{3,2}\rangle|^2\\
 & &+|\langle \phi|A_{3,1}\rangle|B_{3,2}\rangle|^2
 \end{eqnarray*}
is a function over entries of $H$. Even though the analytic expression of the function is complex,
we can compute its maximal value 0.267769 numerically, which is obtained when the system is in the state, written in the
representation of $H$,
 $$H=\left(
      \begin{array}{ccc}
        0.636671 & 0.289003 & 0.197472 \\
        0 & 0.473914 & 0.164544 \\
        0 & 0 & 0.469534 \\
      \end{array}
    \right).$$
To help people reproduce the above results, we list the optimal local measurements in the appendix.

~~In conclusion, there exist two-qutrit and local measurements satisfying the constraint conditions
$P(A_3<B_2)=P(B_2<A_2)=P(A_2<B_1)=P(B_1<A_1)=P(A_1<B_3)=0$ such that the maximal probability of nonlocal event is   $P(A_3<B_3)\approx0.267769.$

  \item $k$=4: Without loss of generality, let Alice and Bob choose VNMs:
$$A_2=B_2=\{|0\rangle\langle 0|,|1\rangle\langle 1|,|2\rangle\langle 2|\},$$
for paradox (\ref{eqH}), we are capable  of  only using entries of $H$ to denote the probability of the nonlocal event $A_4<B_4$, and can
compute its maximal value 0.348158 numerically, which is obtained when the system is in the state
$$H=\left(
      \begin{array}{ccc}
        0.551527 & -0.209186 & 0.184342 \\
        0 & 0.519748 & -0.209186 \\
        0 & 0 & 0.551527 \\
      \end{array}
    \right),
$$
and the optimal local measurements are listed in the appendix.

~~~In short,  there exist two-qutrit and local measurements satisfying the constraint conditions $P(A_4<B_3)=P(B_3<A_3)=P(A_3<B_2)=P(B_2<A_2)=P(A_2<B_1)=P(B_1<A_1)=P(A_1<B_4)=0$ such that the maximal probability of nonlocal event is $P(A_4<B_4)\approx0.348158.$

  \item $k$=5: Let Alice and Bob choose VNMs:
  $$A_3=B_2=\{|0\rangle\langle 0|,|1\rangle\langle 1|,|2\rangle\langle 2|\},$$
  we also have the ability to  obtain the maximal probability of nonlocal event   $P(A_5<B_5)\approx0.40184,$
  which is attained when
  the system's state
  $$H=\left(
      \begin{array}{ccc}
        0.560108 & 0 & 0 \\
        0.17891 & 0.534063 & 0 \\
        0.152703 & 0.17891 & 0.560108 \\
      \end{array}
    \right),$$
and the optimal local measurements, listed in the appendix, satisfy the constraint conditions $P(A_5<B_4)=P(B_4<A_4)=P(A_4<B_3)=P(B_3<A_3)=P(A_3<B_2)=P(B_2<A_2)=P(A_2<B_1)=P(B_1<A_1)=P(A_1<B_5)=0$.
\end{itemize}

The calculations for $k>5$ are beyond our computers'
capability. In table \ref{table2}, we list $SP_{k,3}$ for $k=2,3,4,5$.

\begin{table}[!h]
\caption{\label{table2} For two-qutrit systems, the maximal successful probability to prove nonlocality in Hardy's paradox (\ref{eqH}) for $k=2,3,4,5$.}
\begin{ruledtabular}
\begin{tabular}{ccccc}
 &  $k=2$ &  $k=3$ & $k=4$ & $k=5$\\
  \hline
   $SP_{k,3}$ &  $0.141327$ &$0.267769$ & $0.348158$ & $0.40184$\\
\end{tabular}
\end{ruledtabular}
\end{table}

{\bf Remark 1} For the maximally entangled state (MES), the corresponding matrix $H$ can be written as $\frac{1}{\sqrt{3}}I_3$, and then from the above discussion  one can obtain that   the constrain conditions in (\ref{eqH}) imply $A_k=B_{k-1}=A_{k-1}=B_{k-2}=\cdots=A_1=B_{k}$, which means that MES does not violate the paradox (\ref{eqH}).
Even though the paradox introduced by \cite{Cabello1998} holds for MES, paradox (\ref{eqH}) has more than two times successful probability 0.40184 than 0.171 that of the paradox in \cite{Cabello1998}.

 \section{Generalized Hardy's Inequalities for $k$-settings and $d$-dimensional systems}

Based on the paradox (\ref{eqH}) with $k$-settings and two $d$-dimensional systems, we can have the corresponding  generalized Hardy's inequality as
\begin{widetext}
\begin{equation}\label{eqh4}GH_{k,d}(x,y,z)=\min\{x,y,z\}P(A_k<B_k)-x\sum_{i=2}^kP(A_i<B_{i-1})-
y\sum_{i=2}^kP(B_{i-1}<A_{i-1})-zP(A_1<B_k)\leq 0.\end{equation}
\end{widetext}
with $x>0,y>0,z>0$. Usually for convenience, one can
choose $x, y,z$ as positive integers, and the coefficient $\min\{x,y,z\}$ is used to make the inequality
satisfied by local theory.

\begin{table*}[htbp]
\caption{\label{table4} Maximal values (MV) allowed by the quantum theory (QT) and the maximally entangled states (MES) for   (\ref{eqh4}).}
\begin{ruledtabular}
\begin{tabular}{ccccc}
  &  $k=2$ &  $k=3$ & $k=4$ & $k=5$\\
  \hline
  MV of $GH_{k,3}(1,1,1)$  allowed by QT &  $0.304951$ &$0.429015$ & $0.491126$ & $0.527868$\\
MV of  $GH_{k,3}(1,1,1)$ allowed by MES &  $0.290978$ &$0.414408$ & $0.47795$ & $0.516216$\\
 MV of $GH_{k,3}(2,1,1)$  allowed by QT&  $0.268075$ &$0.393554$ & $0.460468$ & $0.501445$\\
 MV of  $GH_{k,3}(2,1,1)$ allowed by MES &  $0.240055$ &$0.364543$ & $0.434436$ & $0.478489$\\
\end{tabular}
\end{ruledtabular}
\end{table*}

\begin{table*}[htbp]
\caption{\label{table5} Tightness of general Hardy's inequality for $k$-settings and two $d$-dimensional systems.}
\begin{ruledtabular}
\begin{tabular}{c|c|c|c|c|c}
 Tightness of (\ref{eqh4}) &  $k=2$ &  $k=3$ & $k=4$ & $k=5$& $k=6$\\
  \hline
   $d=2$ &  $x=y=z=1,$ Yes &-   & - & - & -\\
   $d=3$ &  - & $x=2,y=z=1,$ Yes & $x=y=z=1,$ Yes & $x=y=z=1,$ Yes& $x=y=z=1,$ Yes\\
   $d=4$ &  - &$x=2,y=z=1,$ Yes & $x=y=z=1,$ Yes & $x=y=z=1,$ Yes & -\\
   $d=5$ &  - & - & $x=y=z=1,$ Yes & - & -\\
\end{tabular}
\end{ruledtabular}
\end{table*}

{\bf Remark 2} In table \ref{table4}, we list the maximally values allowed by quantum theory and
 the maximal entangled states for some  general Hardy's inequalities (\ref{eqh4}).
 Then we find that the optimal states are the nonmaximally entangled states.

{\bf Remark 3}  In table \ref{table5}, we list the tightness of inequalities $GH_{k,d}(x,y,z)\leq 0$, where ``-'' means that we have not found suitable $x,y,z$ such that the inequality $GH_{k,d}(x,y,z)\leq 0$ is tight. Then  we conjecture  that
$GH_{k,2}(1,1,1)\leq 0$ is tight if and only if $k=2$ and for $d=3$ there always exist tight inequalities (\ref{eqh4}) for any $k>2$.

\begin{table}[H]
\caption{\label{table6} The  NTV of the inequalities $GH_{k,2}(1,1,1)$, Hardy inequalities in \cite{Hardy1997} and the chained inequalities.}
\begin{ruledtabular}
\begin{tabular}{cccc}
 &    $k=3$ & $k=4$ & $k=5$\\
  \hline
 NTV of  $GH_{k,2}(1,1,1)$  &    $0.7698$ & $0.811794$ & $0.8411697$\\
 NTV of Hardy inequalities in \cite{Hardy1997}  &    $0.7698$ & $0.811794$ & $0.8411697$\\
 NTV of the chained inequalities   &    $0.7698$ & $0.811794$ & $0.8411697$\\
\end{tabular}
\end{ruledtabular}
\end{table}

{\bf Remark 4}  In table \ref{table6}, we list the nonlocal threshold value (NTV) of the general Hardy's inequalities (\ref{eqh4}) with $x=y=z=1$ and $d=2$, the Hardy's inequalities in \cite{Hardy1997} and the chained Bell inequalities \cite{Pearle1970,Braunstein2}. This means that,
based on the visibility criterion, they are equivalent.

\section{Conclusions}

In this paper, we have presented a general Hardy's paradox for $k$-settings and spin-$s$ systems which generalizes a ladder proof of nonlocality without inequalities and Chen's alternative form of Hardy's paradox. It is well known that improving the success probability to prove nonlocality makes the paradox more adequate for experimental
observation of Hardy-like nonlocality and for applications
based on this type of nonlocality. It is worth to noting that
for spin-$1$ systems, using only $5$-settings, the success probability to prove nonlocality can be improved to 0.4018. Subsequently, we shall try to give the analytic results about the paradox (\ref{eqH}).

\begin{acknowledgments}
H.X.M. is supported by Project funded by China Postdoctoral Science Foundation (No. 2018M631726).
 H.Y.S. is supported by Project funded by China Postdoctoral Science Foundation (No. 2018M630063). T.G. and F.L.Y. are supported by National Natural Science Foundations of China (Grant No. 11475054), Hebei Natural Science Foundation of China (Grant No. A2016205145 and Grant No. A2018205125 ).
 J.L.C. is supported by National Natural Science Foundations of China (Grant No. 11475089).
\end{acknowledgments}

\section*{Appendix}

\subsection{Optimal local measurements in paradox (\ref{eqH}) for $3$-settings and spin-$1$ systems}

$$|A_{1,0}\rangle=|0\rangle,|A_{1,1}\rangle=|1\rangle,|A_{1,2}\rangle=|2\rangle,$$
$$
|A_{2,0}\rangle=\left(
                  \begin{array}{c}
                     0.840759 \\
                     -0.512713 \\
                     -0.173922 \\
                  \end{array}
                \right)
,|A_{2,1}\rangle=\left(
                   \begin{array}{c}
                      0.39627 \\
                      0.801645 \\
                       -0.447589 \\
                   \end{array}
                 \right),$$$$
  |A_{2,2}\rangle=\left(
                   \begin{array}{c}
                      0.368908 \\
                     0.307394 \\
                       0.877163 \\
                   \end{array}
                 \right),
|A_{3,0}\rangle=\left(
                  \begin{array}{c}
                      0.604857 \\
                      -0.783492 \\
                      -0.142437 \\
                  \end{array}
                \right)
,$$
$$|A_{3,1}\rangle=\left(
                   \begin{array}{c}
                      -0.361768 \\
                       -0.429694 \\
                      0.827337 \\
                   \end{array}
                 \right),
  |A_{3,2}\rangle=\left(
                   \begin{array}{c}
                       0.709416 \\
                      0.448892 \\
                       0.543346 \\
                   \end{array}
                 \right), $$
$$|B_{1,0}\rangle=|0\rangle,|B_{1,1}\rangle=|1\rangle,|B_{1,2}\rangle=|2\rangle,$$
$$
|B_{2,0}\rangle=\left(
                  \begin{array}{c}
                     -0.0700471 \\
                     0.0357429 \\
                     0.0138888 \\
                  \end{array}
                \right)
,|B_{2,1}\rangle=\left(
                  \begin{array}{c}
                    0.00836581 \\
                    0.021817 \\
                    -0.0139536 \\
                  \end{array}
                \right),$$
$$|B_{2,2}\rangle=\left(
                  \begin{array}{c}
                    0.125725 \\
                    0.13505 \\
                     0.286533 \\
                  \end{array}
                \right),|B_{3,0}\rangle=\left(
                    \begin{array}{c}
                       0.876299\\
                       0.397777 \\
                       0.271796 \\
                    \end{array}
                  \right)
,$$
$$|B_{3,1}\rangle=\left(
                    \begin{array}{c}
                      -0.460165 \\
                       0.858127 \\
                       0.227742 \\
                    \end{array}
                  \right),|B_{3,2}\rangle=\left(
                    \begin{array}{c}
                       0.142645 \\
                       0.324641 \\
                       -0.935019 \\
                    \end{array}
                  \right).$$

\subsection{Optimal local measurements in paradox (\ref{eqH}) for $4$-settings and spin-$1$ systems}

$$|A_{1,0}\rangle=\left(
                    \begin{array}{c}
                      0.905512 \\
                      -0.349204 \\
                      0.241051 \\
                    \end{array}
                  \right)
,|A_{1,1}\rangle=\left(
                    \begin{array}{c}
                      -0.399076 \\
                      -0.893898 \\
                      0.204168 \\
                    \end{array}
                  \right),$$
$$|A_{1,2}\rangle=\left(
                    \begin{array}{c}
                      0.144179 \\
                      -0.281074 \\
                      -0.948794 \\
                    \end{array}
                  \right),$$
$$|A_{2,0}\rangle=|0\rangle,
                  |A_{2,1}\rangle=|1\rangle,|A_{2,2}\rangle=|2\rangle,
$$
$$|A_{3,0}\rangle=\left(
                    \begin{array}{c}
                       0.914794 \\
                       0.368182 \\
                       -0.166114 \\
                    \end{array}
                  \right)
,|A_{3,1}\rangle=\left(
                    \begin{array}{c}
                      -0.272352 \\
                      0.865949 \\
                      0.419472 \\
                    \end{array}
                  \right),$$
$$
  |A_{3,2}\rangle=\left(
                    \begin{array}{c}
                      0.298288 \\
                      -0.338489 \\
                      0.89244 \\
                    \end{array}
                  \right),|A_{4,0}\rangle=\left(
                    \begin{array}{c}
                       0.800697 \\
                       0.573965 \\
                       -0.171602 \\
                    \end{array}
                  \right),$$
$$|A_{4,1}\rangle=\left(
                    \begin{array}{c}
                       0.307453 \\
                       -0.639559 \\
                       -0.704583 \\
                    \end{array}
                  \right),|A_{4,2}\rangle=\left(
                    \begin{array}{c}
                       0.514156\\
                       -0.511398\\
                       0.68856\\
                    \end{array}
                  \right),$$
  $$|B_{1,0}\rangle=\left(
                    \begin{array}{c}
                       0.89244 \\
                       -0.338489 \\
                       0.298288 \\
                    \end{array}
                  \right)
,|B_{1,1}\rangle=\left(
                    \begin{array}{c}
                      0.419472 \\
                       0.865949 \\
                       -0.272352 \\
                    \end{array}
                  \right),$$
$$|B_{1,2}\rangle=\left(
                    \begin{array}{c}
                       0.166114\\
                       -0.368182 \\
                       -0.914794 \\
                    \end{array}
                  \right)
,$$
$$|B_{2,0}\rangle=|0\rangle,|B_{2,1}\rangle=|1\rangle,|B_{2,2}\rangle=|2\rangle,$$
$$|B_{3,0}\rangle=\left(
                    \begin{array}{c}
                       -0.948794\\
                      -0.281074 \\
                      0.144179 \\
                    \end{array}
                  \right)
,|B_{3,1}\rangle=\left(
                    \begin{array}{c}
                      -0.204168 \\
                       0.893898 \\
                      0.399076 \\
                    \end{array}
                  \right),$$
$$|B_{3,2}\rangle=\left(
                    \begin{array}{c}
                       0.241051 \\
                       -0.349204 \\
                       0.905512 \\
                    \end{array}
                  \right),|B_{4,0}\rangle=\left(
                    \begin{array}{c}
                       0.68856\\
                       -0.511398\\
                       0.514156\\
                    \end{array}
                  \right),$$
$$|B_{4,1}\rangle=\left(
                    \begin{array}{c}
                       -0.704583\\
                       -0.639559\\
                      0.307453 \\
                    \end{array}
                  \right),|B_{4,2}\rangle=\left(
                    \begin{array}{c}
                       0.171602\\
                      -0.573965 \\
                      -0.800697 \\
                    \end{array}
                  \right).$$

\subsection{Optimal local measurements in paradox (\ref{eqH}) for $5$-settings and spin-$1$ systems}

$$|A_{1,0}\rangle=\left(
                    \begin{array}{c}
                       0.749824\\
                       0.477546 \\
                       0.457945 \\
                    \end{array}
                  \right),|A_{1,1}\rangle=\left(
                    \begin{array}{c}
                       0.634354\\
                       -0.715587\\
                       -0.292456\\
                    \end{array}
                  \right),$$
$$|A_{1,2}\rangle=\left(
                    \begin{array}{c}
                       0.188038\\
                       0.50979\\
                       -0.839497\\
                    \end{array}
                  \right),|A_{2,0}\rangle=\left(
                    \begin{array}{c}
                       0.921999\\
                       0.294506\\
                       0.251365\\
                    \end{array}
                  \right),$$
$$|A_{2,1}\rangle=\left(
                    \begin{array}{c}
                       -0.354751\\
                       0.902659\\
                       0.243634\\
                    \end{array}
                  \right),|A_{2,2}\rangle=\left(
                    \begin{array}{c}
                      0.155145 \\
                      0.313803 \\
                      -0.936727 \\
                    \end{array}
                  \right),
$$
$$|A_{3,0}\rangle=|0\rangle,|A_{3,1}\rangle=|1\rangle,|A_{3,2}\rangle=|2\rangle,
$$
$$|A_{4,0}\rangle=\left(
                    \begin{array}{c}
                       -0.958481\\
                       0.248819\\
                       0.139295\\
                    \end{array}
                  \right),|A_{4,1}\rangle=\left(
                    \begin{array}{c}
                       0.187595\\
                       0.918099\\
                       -0.349146\\
                    \end{array}
                  \right),$$
$$|A_{4,2}\rangle=\left(
                    \begin{array}{c}
                       0.214761\\
                       0.308518\\
                      0.926658 \\
                    \end{array}
                  \right),
|A_{5,0}\rangle=\left(
                    \begin{array}{c}
                       0.898853 \\
                       -0.40709 \\
                       -0.162297 \\
                    \end{array}
                  \right),$$
$$|A_{5,1}\rangle=\left(
                    \begin{array}{c}
                      -0.251302 \\
                       -0.782172 \\
                       0.570136 \\
                    \end{array}
                  \right),|A_{5,2}\rangle=\left(
                    \begin{array}{c}
                       0.35904 \\
                       0.471683 \\
                       0.80536 \\
                    \end{array}
                  \right),
$$
$$|B_{1,0}\rangle=\left(
                    \begin{array}{c}
                       0.926658\\
                       0.308518\\
                       0.214761\\
                    \end{array}
                  \right),|B_{1,1}\rangle=\left(
                    \begin{array}{c}
                       0.349146\\
                       -0.918099\\
                       -0.187595\\
                    \end{array}
                  \right),$$
$$|B_{1,2}\rangle=\left(
                    \begin{array}{c}
                       0.139295\\
                       0.248819\\
                      -0.958481 \\
                    \end{array}
                  \right),
$$
$$|B_{2,0}\rangle=|0\rangle,|B_{2,1}\rangle=|1\rangle,|B_{2,2}\rangle=|2\rangle,
$$
$$|B_{3,0}\rangle=\left(
                    \begin{array}{c}
                       0.936727\\
                       -0.313803\\
                       -0.155145\\
                    \end{array}
                  \right),|B_{3,1}\rangle=\left(
                    \begin{array}{c}
                       0.243634\\
                       0.902659\\
                       -0.354751\\
                    \end{array}
                  \right),$$
$$|B_{3,2}\rangle=\left(
                    \begin{array}{c}
                       0.251365\\
                       0.294506\\
                      0.921999 \\
                    \end{array}
                  \right),|B_{4,0}\rangle=\left(
                    \begin{array}{c}
                      0.839497 \\
                      -0.50979 \\
                      -0.188038 \\
                    \end{array}
                  \right),$$
$$|B_{4,1}\rangle=\left(
                    \begin{array}{c}
                      -0.292456 \\
                      -0.715587 \\
                      0.634354 \\
                    \end{array}
                  \right),|B_{4,2}\rangle=\left(
                    \begin{array}{c}
                       0.457945\\
                       0.477546\\
                       0.749824\\
                    \end{array}
                  \right),
$$
$$|B_{5,0}\rangle=\left(
                    \begin{array}{c}
                      0.80536 \\
                       0.471683 \\
                       0.35904 \\
                    \end{array}
                  \right),|B_{5,1}\rangle=\left(
                    \begin{array}{c}
                       0.570136 \\
                       -0.782172\\
                       -0.251302\\
                    \end{array}
                  \right),$$
$$|B_{5,2}\rangle=\left(
                    \begin{array}{c}
                      0.162297 \\
                      0.40709 \\
                      -0.898853 \\
                    \end{array}
                  \right).
$$

\end{document}